\documentclass[review]{elsarticle}

\usepackage{lineno,hyperref}

\usepackage{amsmath}
\usepackage{amssymb} 
\graphicspath{{figures/}}

\newcommand{\arcsec}{\hbox{$^{\prime\prime}$}}

\newcommand{\farcs}{\hbox{$.\!\!^{\prime\prime}$}}

\newcommand{\muas}{\hbox{$\mu$}as}

\def\cent{\multicolumn{1}{c}}

\modulolinenumbers[5]

\journal{Journal of \LaTeX\ Templates}

\begin{document}

\begin{frontmatter}

\title{Astronomical distance scales in the Gaia era}

\author{F. Mignard\fnref{myfootnote}}
\address{Universit\'{e} C\^{o}te d'Azur, Observatoire de la C\^{o}te d'Azur, CNRS, Laboratoire Lagrange}
\address{Bd de l'Observatoire, CS 34229, 06304 Nice Cedex 4, France}
\ead{francois.mignard@oca.eu}
\fntext[myfootnote]{Published in Comptes Rendus Physique, Special issue: The new International System of Units / Le nouveau Système international d'unités \url{https://www.sciencedirect.com/science/article/pii/S1631070519300155} }

\begin{abstract}
Overview of the determination of astronomical distances from a metrological standpoint. Distances are considered from the Solar System (planetary distances) to extragalactic distances, with a special emphasis on the fundamental step of the trigonometric stellar distances and the giant leap recently experienced in this field thanks to the ESA space astrometry missions Hipparcos and Gaia.
\end{abstract}

\end{frontmatter}


\section{Introduction}
For  centuries astronomers had to content themselves with a 2-dimensional world with virtually no access to the depth of the Universe. The world unfolded before their eyes as though everything was taking place on the surface of a spherical envelope with few exceptions for the nearest sources, such as the Moon whose nearness was made obvious from its repeated passages before the Sun (solar eclipses), the planets or the stars (occultations). The size of this sphere was arbitrary and could not be gauged, let alone the idea that the stars could lie at different distances.  Until the 17th century a reliable estimate of the true distance to the Sun and of the  size of the Solar System remained out of reach, although a good scale model could be accurately devised and actually crafted in the form of delicately adorned orreries (but not all were on-scale). 

Regarding the sidereal world and the immense vacuum lying beyond Saturn before reaching the first stars, some realistic ideas started emerging a good century later with the assumption that stars are Suns and share more or less the same luminosity. Gregory, Huygens among others came to numbers that at least hinted at the immensity of the world lying beyond the solar system. However the first indisputable stellar distances free of any physical assumption about the nature of the stars came out in 1840 through three independent labours, among which that of F.W. Bessel stands out. Once this first direct step has been mastered astronomers developed gradually a whole set of methods to ascertain the distances of celestial objects, each new step going farther in the cosmos and depending on the reliability of the previous rungs.

This short review aims at an audience of scientists with no particular astronomical background beyond the general knowledge shared by every physicist. Simple and basic formulas that would not appear in an astronomical research paper are given and explained. Only the principles of the methods are provided, illustrated on simple cases, leaving out the real difficulties which are the daily bread of practitioners.  The book \cite{1985cdld.book.....R} by M. Rowan-Robinson provides a more technical and comprehensive review of the subject from stars to cosmological distances. Published before Hipparcos and HST, the content is a bit outdated but the description of the issues and the astronomical principles are still valuable and could be complemented with the more recent review of S. Webb \cite{1999meun.book.....W}. At the solar system level the monograph \cite{1986mucd.book.....V} by A. van Helden is the best reference for the historical coverage from Aristarchus to Halley, but includes nothing relevant for the modern period.

The text is organised in two major sections. The first deals with distances within the solar system with the length of the astronomical unit in kilometres to its recent conceptual mutation to a defining constant with a fixed relation to the SI unit of length. The second part covers the scale of the Universe from the stars to the cosmological distances, with a particular emphasis on the first fundamental rung of the ladder completely rejuvenated over the last twenty years with the two ESA astrometry satellites Hipparcos and Gaia.

\section{Distances in the Solar System: the astronomical unit}
\subsection{Relative vs. absolute sizes in the solar system}
Astronomical distances have practically never been measured or numerically expressed with standard metric units, like \textsl{m} or \textsl{km}. First this would not be convenient units given the size of the solar system, let alone the distances of the stars or that of the galaxies. One could claim with good reasons that this can be resolved by a proper choice of multiples, and this will not put astronomy aside from the SI system. This is true  and there is a more fundamental ground for the use of an independent and consistent system of units in astronomy. 

Except in very limited and relatively recent instances with radar and laser ranging in the solar system, measured space quantities in astronomy are always angles and not lengths or distances as it is on the Earth. Therefore distances are derived quantities and byproducts of astrometric measurements attempting to detect small angular shifts in the direction of a celestial body resulting from its observation from at least two different points, as distant as possible from each others. The baselines, the Earth's radius or the size of the orbit of the Earth around the Sun, were not necessarily known in metric units with an accuracy matching that permitted with the angular measurements. This issue is more important in the solar system than it is for the stars and the galaxies, for which no extreme fractional error is achievable, even today with Gaia, the on-going ESA Astrometry mission, or the HST (Hubble Space Telescope), the only providers of direct and accurate measurements of stellar distances in the visible, although radio astronomy can do even better on a small number of galactic H2O or OH masers \cite{2017ApJ...849...99Z}. 

In the solar  system the relative size of the planetary orbits was known to a good accuracy even before the discovery of the third Kepler's law, relating the orbital period to the distance to the Sun. From pure angular observations it was possible at the time of Copernicus to build a model of the solar system showing the orbit of Mars or Venus with their correct scale compared to the Earth with a precision of about $5\%$. However the absolute scale expressed either in Earth radii, feet or toises was not possible without loss of accuracy. This situation worsened, in some sense, when orbits could be computed with the laws of gravitation as the relative accuracy greatly improved and the gap between the relative and absolute size widened. The need to use a reference of length disconnected from the standards used for trade or scientific usage became mandatory to benefit fully from the accurate astrometry.

\subsection{The astronomical system of units}
Starting in the 19th century and made official by IAU in 1938, the astronomical unit was defined as a fundamental constant of the astronomical system of units as a length such that the Gravitational constant is the square of the defining Gauss constant, 
\begin{equation} \label{gauss_cte}
k   = 0.017,202,098,95
\end{equation}
yielding,
\begin{equation}\label{newton_cte}
G = k^2 = 0.000,295,912,208,285,591,102,5
\end{equation}
with the unit of mass being the solar mass and the unit of time the solar day of $86,400$ seconds. Combined with the Kepler's third law,
\begin{equation}\label{kepler}
\frac{a^3}{P^2}  = \frac{GM_\odot}{4\pi^2}
\end{equation}
Eq.~\ref{gauss_cte}, implies that the mean motion of a massless planet orbiting the Sun at one astronomical unit is $k$ rad\,day$^{-1}$, corresponding to a period of 
\begin{equation} \label{period}
P   = \frac{2\pi}{k} = 365.25689832 \text{ days}
\end{equation}
very close to the sidereal year.
Therefore the $au$ defined by Eqs.~\ref{gauss_cte}-\ref{kepler} agrees with the simple initial idea of the astronomical unit being essentially the mean distance between the Earth of the Sun, or  the semi-major axis of its orbit, although this is not formally its definition. With the above definition and units, the law of attraction reads,
\begin{equation}\label{Nlaw_k}
\frac{d^2\mathbf{r}}{dt^2}  = -\frac{k^2\mathbf{r}}{r^3}
\end{equation}
This allowed astronomers to produce very accurate numerical or analytical theories of the motion of solar system bodies and predict their  positions  without having their absolute distances. The whole system is consistent and angular observations constrain the free constants of the model, primarily the position and velocity vectors of the bodies at an arbitrary epoch.

\begin{figure}[h]
  \begin{center}
   \includegraphics[width=10cm]{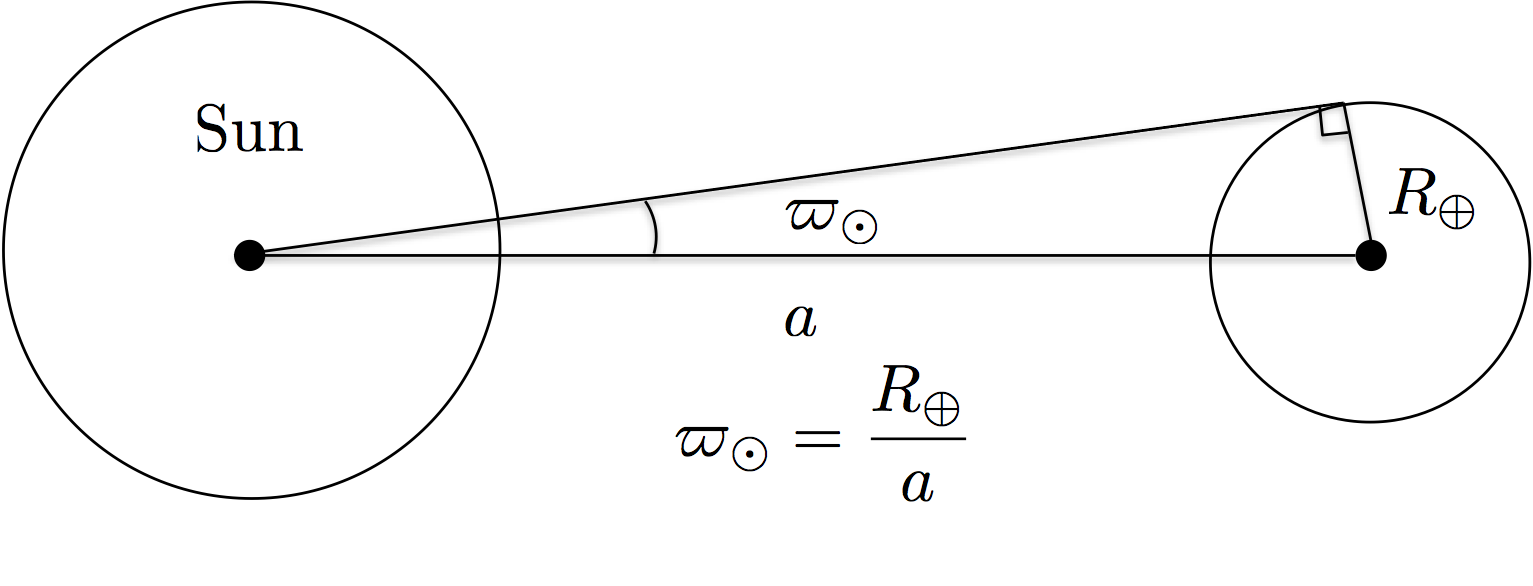}
  \end{center}
  \vspace{-5mm}
\caption{Definition of the solar parallax}
\label{Fig:solar_varpi}
\end{figure}

\subsection{The solar parallax}
The absolute length of the $au$  was derived from dedicated observations as an angular quantity called the solar parallax whose meaning is shown in Fig.~\ref{Fig:solar_varpi}. The measurement of the angle $\varpi_\odot$  being equivalent to the Sun distance expressed with the Earth's radius. The latter being known in common unit, the procedure would yield the size of the solar system in the same unit. Assessing this length remained a central issue in astronomy until very recently and was even referred to as  \emph{the noblest problem in astronomy} by G. B. Airy, the Astronomer Royal from 1835 to 1881. With the law of motion, a single measurement of a distance of one solar system body suffices in establishing the absolute scale of the solar system. The very first significant step in this direction was achieved in 1672 during a most favourable (small distance) opposition of Mars, making its parallax as large as possible. On the 04 September 1672, Mars approached the Earth at $0.381 $ au, very close to the smallest possible distance of $0.371$ au.  Observing from two distant points J.D Cassini in Paris and J. Richer in Cayenne found an equatorial parallax of $10$\arcsec, or

\begin{equation}
\frac{R_{\oplus}}{\text{1 au}}  = 10\arcsec
\end{equation}

giving, $1 \text{au} \approx 1.32\times 10^8 \text{ km}$. This is too small by $\approx 10\%$, but for the first time astronomers had a sensible estimate of the real size of the solar system from a method whose principle was sound and could not be disputed.

The rare transits of Venus across the solar disk offered another way of ascertaining the au as noted first by J. Gregory in 1663 and widely heralded by E.~Halley in 1716. The advantage of Halley's proposal is still extant since he proposed to replace pure angular measurements by timings of the moment the dark disk of the planet is seen encroaching on the bright solar disk. Given the angular speed of Venus relative to the Sun it is easy to show that a better accuracy can in principle be reached with the timing than with classical position sights. Halley claimed that the transit duration could be assessed to few seconds of time and consequently the distance to the Sun to one part to few thousandths. International cooperation was put in place for every following occurrence of the Venus transit in 1761, 1769, 1874, 1882 to observe and time the passages from the most remote places on the Earth. This led to adventurous expeditions that have been reported in many books and most is available on-line or in popular accounts \cite{1959tvse.book.....W}, \cite{2000vetr.book.....M}. 

Regarding the astronomical aim, the results were not on a par with the expectations and never reached the accuracy claimed by the illustrious astronomer. The extensive discussion of the four transits by S. Newcomb in 1892 ended up with a solar parallax of $ \varpi_\odot = 8\farcs 79 \pm 0 \farcs 018$  (current determination $8 \farcs 794143\cdots$) or a value for the Sun-Earth distance of $(149.7 \pm 0.3)\times 10^6$ km. This was in some sense a very unsatisfactory situation in regards of the achievements of the planetary theories at the same time and after the triumph of the solar system dynamics with the discovery of Neptune in 1846. 

A fortunate circumstance cast some lights in a gloomy landscape with the discovery in 1898 of the minor planet Eros (433 Eros) simultaneously at Berlin and Nice, the first of the near-Earth objects to be identified. Eros comes within the orbit of Mars and favourable oppositions that repeat every 30 years may bring the planet to $0.2$ au \footnote{The astronomical unit should be abbreviated as \textsl{au} since 2012 as stated in the recommendation of the International Astronomical Union in its Resolution B2. This is also the notation given by the BIPM in its official list of secondary units. It is usual to find instead AU or ua.}  from the Earth, closer than any other solar system object known at that time. The first such passage took place in 1901 and the next good one was in 1931. Again a broad international cooperation was set up to observe and reduce the observations and led to a solar parallax of $ \varpi_\odot =8\farcs 790 \pm 0 \farcs 001$. It was the most accurately known value for the solar distance at that time, and this value has remained the standard until mid-1960 when radar measurements gave a more accurate value for the distance to the Sun. 

\subsection{The astronomical unit today}
Again a direct range measurement based on timing took precedence over classical angular measurements, with a measured  quantity that was almost a distance, and no longer an angle. In particular there were no more reasons to express it as a \textsl{parallax}, a formulation inherited from the measurement technique, but a distance expressed directly in SI units, given the accuracy of the velocity of light. The distance became the primary quantity and the parallax  a derived parameter. Later on the use of spacecraft tracking combined to highly accurate global numerical integrations of the solar system motions resulted into the best values of the astronomical unit (\cite {2009CeMDA.103..365P}, \cite {2008A&A...477..315F}, \cite{2011CeMDA.111..363F}), which eventually led the International Astronomical Union to recommend in 2009 (Resolution B2, IAU 2009 System of Units) a value of $ 149,597,870,700 \pm 3$ m for the au.

Eventually this was turned into a defining astronomical constant in the IAU 2012 Resolution B2 with the  astronomical  unit  being  a  conventional  unit  of  length  strictly equal  to $ 149,597,870,700$\,m in agreement with the value adopted in IAU 2009 Resolution B2 \cite{2012IAUJD...7E..40C}. Accordingly the BIPM changed  this unit from the table of non-SI units whose values in SI units must be obtained experimentally to the table of  non-SI units accepted for use with the International System of Units. It is now tied to the meter with a fixed factor. In short this is now a multiple of the meter and what should be  experimentally determined is the scale factor of the solar system, say the Sun-Earth mean distance expressed in au. A consequence is that to the equation of motion (\ref{Nlaw_k}) one must substitute,
\begin{equation}\label{Nlaw_SI}
\frac{d^2\mathbf{r}}{dt^2}  = -\frac{GM_\odot\mathbf{r}}{r^3}
\end{equation}
with $GM_\odot$ in m$^3$\,s$^{-2}$ and the SI units or their multiples for length and time. Modern numerical integrations of the Solar System comply now with this requirement.
As far as metrology is concerned the situation is clarified and it is left to the astronomers now to refine their measurements to give the size of the orbits in meters with the best accuracy.

\section{Distance of the stars}
 \subsection{The trigonometric parallaxes}
\begin{figure}[h]
  \begin{center}
   \includegraphics[width=10cm]{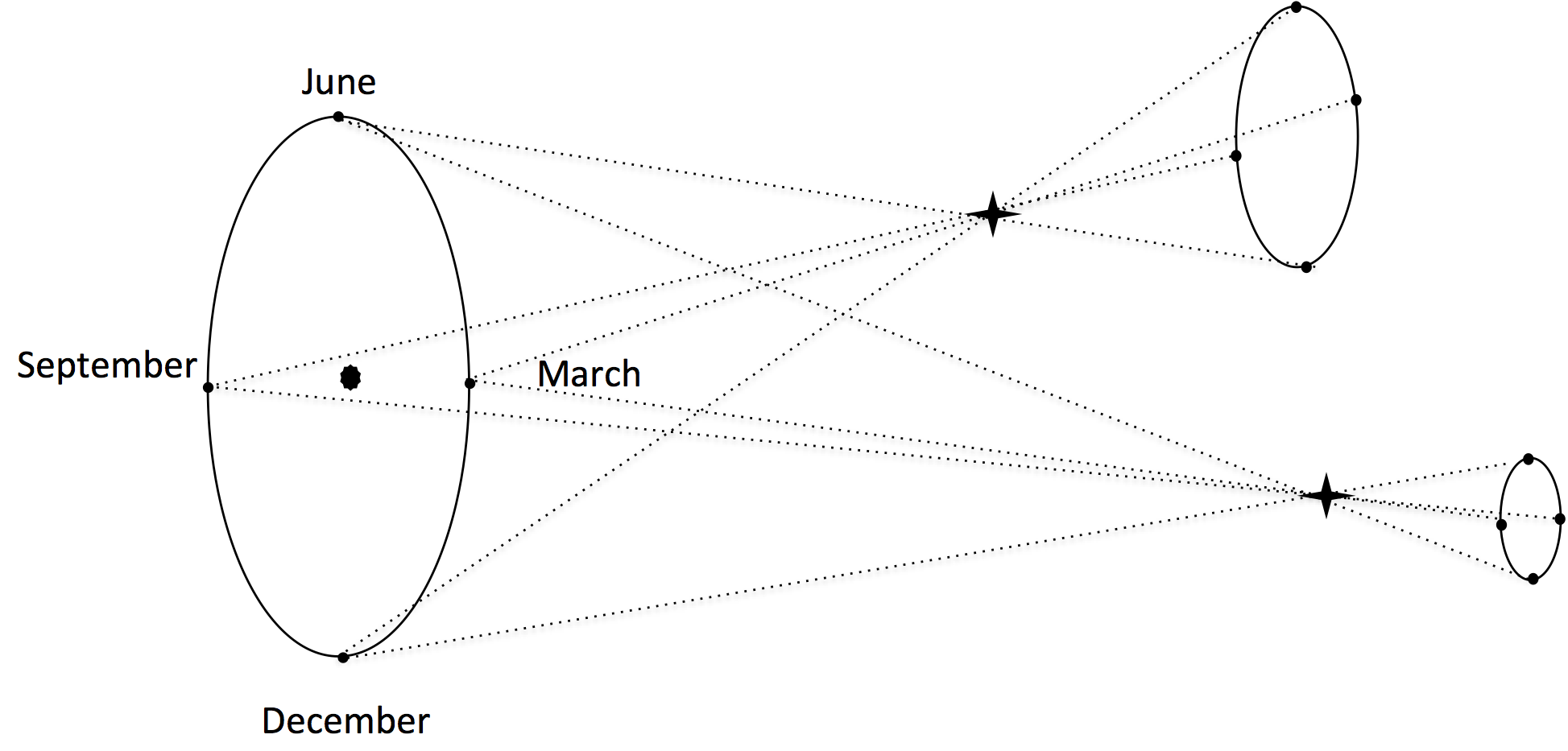}
  \end{center}
  \vspace{-5mm}
\caption{The parallactic motion for a nearby and a distant star.}
\label{Fig:annual_varpi}
\end{figure}

For centuries the problem of stellar distances has puzzled astronomers, although the underlying geometric principles needed to ascertain them  were extremely simple and well understood. The basic idea is sketched out in Fig.~\ref{Fig:annual_varpi} showing the apparent shift in the star position resulting from the annual motion of the Earth around the Sun.

Provided the stars are not infinitely remote compared to the size of the Earth's orbit, our annual displacement  translates into a reflex apparent displacement of the stars on the sky, since during the year the different lines joining the observer to the star are not parallel. The farther the star, the smaller the parallactic ellipse, and more precisely its size is proportional to the reciprocal of the star distance. The parallax of a star is  defined by the angle subtended at the star by one astronomical unit or half the apparent diameter of the Earth orbit when seen from the star. Mathematically one has for the  parallax $\varpi$ of a star at distance $d$ from the solar system,
\begin{equation}\label{def_varpi}
\varpi =  \frac{a}{d}
\end{equation}
where $a =1$ au. The unit of distance is the \textsl{parsec}, noted pc and  its common multiples are kpc, Mpc, Gpc. By definition of the parsec, a star at 1pc has a parallax of $1 \farcs 0$, meaning that its distance is $206,264$ au, corresponding also to $3.26$ light-years  or $3.1\times 10^{16}$ km. With distances in pc and parallaxes in arcsec one has $ \varpi = 1/d$. No star has a parallax reaching one arcsec and the nearest star, Proxima Centauri with a parallax of $0\farcs 77$ is at 1.30 pc from us. In astronomy one does not measure distances directly, but their angular signature with the parallax. Hipparcos and Gaia deliver parallaxes from which distances are inferred. But the parallax is the best way to ascertain distances within the Galaxy without any assumption regarding the physics of the source, although this is not the only mean to do it. However this is the only way to do it on large number of stars in a survey mode, while the alternative method needs eclipsing binary stars with good spectroscopy, a relatively rare instance.

Given the definition of this unit of distance, it is clear that both the measured parallaxes and the distances expressed in parsec  are independent of the precise knowledge of the au in metric units and would not change with an improvement of the au. This is a direct expression of the angular shift on the sky with a distance given in Sun-Earth distance. Using lightyear would have been an option, but nothing in the measurement principle would allow to link naturally the parallax to the light travel time and simply to the finite speed of the propagation of light. The choice of the pc has been discussed around 1910 and made official by the IAU in 1925. Within the Galaxy distances are given in pc or kpc, while extragalactic distances are in Mpc. Beyond few Gpc the redshift, which is the measured quantity, is more usual. Also the distance modulus  given in Eq.~\ref{m_minus_M} is convenient when photometric data are involved. The diameter of the Milky Way is about $\approx 30$ kpc and the Sun is located at $\approx 8$ kpc from the galactic centre. The Andromeda galaxy, our nearest external major galaxy, lies at $0.78$ Mpc and has a distance modulus of $24.5$ mag.

It is also important to stress that in the heliocentric theory detecting the parallactic motion is a proof of the Copernican doctrine, and conversely its opponents exploited the lack of detection to support alternative theories and to challenge the doctrine. Therefore the signature of the Earth's motion was primarily searched for fundamental reasons  rather than to learn about the size of the Universe. The Tycho planetary system was a partial answer to this absence of evidence and could not be opposed as long the parallax of the fixed stars, or any other proof of the Earth motion, was not seen. This came before the first stellar parallax was measured through the discovery of the stellar aberration in 1727 by J. Bradley, while he was himself engaged into the search of the stellar parallax.

The tiny parallactic motion shows up as a periodic change of spherical coordinates as, 
\begin{align}\label{equat_varpi}
\Delta\alpha \cos\delta &=  -\varpi r(\sin\alpha\cos\lambda_\odot- \cos\alpha\sin\lambda_\odot\cos\epsilon )\\ 
\Delta \delta\phantom{\cos\delta} &= -\varpi r \left[\sin\lambda_\odot(\cos\epsilon \sin\delta\sin\alpha - \sin\epsilon \cos\delta)+\cos\lambda_\odot\sin\delta\cos\alpha  \right]
\end{align}
respectively for the  star right-ascension and declination. Here, $\varpi$ is the parallax (usually given in second or millisecond of degree), $\alpha, \delta$ are the right-ascension and declination, $\lambda_\odot$ the ecliptic longitude of the Sun, $\epsilon$ the obliquity of the ecliptic and $r$  the distance of the observer to the Sun in astronomical unit (always very close to unity, even with Gaia).  The path described by the star on the sky is an ellipse of semi-axes $\varpi,  \varpi \sin \beta$, where $\beta$ is the ecliptic latitude. This is a circle at the ecliptic pole which degenerates into a straight segment of length $2\varpi$  in the ecliptic plane.

In an ideal world, parallaxes could be found by sampling the equatorial coordinates of a star over a year and then extracting the amplitude of the yearly sinusoidal change in one or both coordinates. But the amplitude is at most $0 \farcs 7$ in the most favourable case and two to three orders of magnitude smaller for a typical galactic star. In addition there are other sources of change in the star coordinates which must be accounted for and the parallax is usually a small fraction of the whole motion. Getting accurate absolute parallaxes is nearly hopeless with ground-based observations given the adverse effect of the refraction, the telescope flexure and the difficulty to refer observations to an invariable frame of reference during the year. 

As noted by Galileo resorting to a small field offered a route to success. Instead of measuring the absolute displacement in a well defined reference frame, one could detect the tiny parallactic motion with respect to one or few neighbouring stars with the additional assumption that these reference stars are far more distant than the star whose parallax is searched. In short the measurement is no longer $\varpi$ but the difference between the parallax of the nearby star and that of the reference star(s). One gets at the end a relative parallax instead of an absolute parallax, as long as one cannot tell how far is the reference star to correct the result for this bias. After many failed attempts, the German astronomer and mathematician F.W. Bessel made the first successful parallax measurement ever, for the star 61 Cygni which he found equal to $0\farcs 314$ ($0\farcs 285 \pm0\farcs 0005 $ with Hipparcos and  $0\farcs 28615 \pm0\farcs 00006 $ with Gaia for 61 Cygni B), or a distance of $3.2$ pc (10.3 lightyears).

This marked the start of a systematic and difficult search which is still on-going today with better instruments placed outside the Earth's atmosphere.  This allowed astronomers to get about 100 measured parallaxes by the year 1900 with a relative accuracy better than 50\%. The number grew steadily during the 20th century, as shown in Table~\ref{table:stat_varpi}, but this remained a painstaking task with low yielding, although the use of photographic plates from $\approx 1920$ onwards relieved observers from long hours at the eyepiece in the near open air, traded for equivalent long hours at the measuring machine in the comfort of a laboratory. 
\begin{table}[t]
\centering
        \caption{Number of the published stellar parallaxes (number of stars with at least one parallax) before Hipparcos. \label{table:stat_varpi}}    
        \begin{tabular}{rrr}
        \noalign{\medskip  }
          \hline\hline
        \noalign{\smallskip}
        Year  &\cent{number}  &\cent{notes} \\[2pt]
         \hline         
        \noalign{\smallskip} 
        1840	&3 & 61 Cygni, Vega, $\alpha$ Cent\\
        1850    & 20 & \\
        1890    & 40 & \\
        1910    & 300& with 52 photographic parallaxes \\
        1925    & 2000 & photographic plates \\
        1965    & 6000 & Yale catalogue \\
        1980    & 8000 & just before Hipparcos\\
        \hline
        \end{tabular}
\end{table}
See \cite{1909BuAsI..26..291B} for a discussion of the state of the art around 1910.  

The large scatter of the measurements carried out on the same stars by different observers and different methods gives an idea of the systematic errors. The fourth, and last, version of the Yale Catalogue of Trigonometric Parallaxes \cite{2001yCat.1238....0V}, the reference in the field before Hipparcos, gives the trigonometric parallaxes for $8112$ stars with a mode in the quoted accuracy of $0\farcs 004$, an improvement of a factor three compared to the previous release of 1963. To appreciate the  difficulty to obtain parallaxes with ground-based astrometric techniques, in the interval of 32 years between the two publications only $1,722$ stars were added.

 In any case the number of reliable trigonometric parallaxes, say better than 10\% in relative uncertainty, stayed below $\approx 2000$ before the launch of Hipparcos and remained limited to bright stars. This was really very small in comparison with the contemporary sky surveys listing several 100's million of stars with positions and some basic photometric information. Of course the trigonometric parallaxes were not the only distance estimator available, but this was the only way to get a geometric measurement of the parallax free of any assumption on the physics of the stars, and any other method had to be calibrated on reliable distances and ultimately rested upon this small set of trigonometric parallaxes. 
     
 \subsection{The Hipparcos parallax survey}\label{sect:hipparcos}
Hipparcos (see for reviews \cite{2012EPJH...37..745P}, \cite{2011A&ARv..19...45P}) opened a new era for astrometry thanks to the access to space to do accurate astrometric observations without the limitations caused by the bending and twinkling of light rays by the Earth's atmosphere. The key ideas to carry out absolute astrometric measurements from space can be traced back to the French Astronomer Pierre Lacroute in the mid-1960s. He realised that a census could be carried out with almost uninterrupted observations allowing to triangulate the heavens with long arcs between pairs of stars. The most important and novel idea was to observe simultaneously two fields of view in very widely separated directions. Combining these two lines of sight onto a single focal plane would lead to a rigid network of stellar positions covering the whole celestial sphere, provided one could guarantee the stability of the angle between the two directions.

This was still basically relative astrometry, but not the way astronomers were used to, thanks to the wide angular separation between the two directions. The phase of the parallactic ellipses in each direction would be different and stars linked by the wide angle would change from time to time. At the end one could reconstruct the 2-D position of each star at a reference epoch and  solve for their motion and absolute parallax, at the expense of a global adjustment  incorporating all the observations. After a long process of maturation, feasibility assessment, technical studies and lobbying, the Hipparcos project was selected in the European Space Agency's science program in July 1980 and eventually the satellite was launched in August 1989. The scientific goal was to reach a 2 milliarcsec (herein abbreviated as \textsl{mas} for $0\farcs 001$) accuracy for about 100,000 stars, selected on the basis of their astrophysical interest, survey coverage at around 8.5 mag and ability of Hipparcos to observe them repeatedly during its planned mission of two years. The launch was dramatic and near fatal to the mission due to the failure to fire the apogee motor needed to reach the geostationary orbit. This ended up with the satellite on a wrong orbit, from which virtually nothing valuable was believed achievable and a life-expectancy much reduced due to the daily crossing of radiation belts. Eventually the mission scenario was adapted and successful operations lasted until march 1993 when to much accumulation of hig-energy particles fatally impaired the electronic. The final results published by ESA in June 1997 \cite{1997A&A...323L..49P}  surpassed the expectations placed on the satellite at its acceptance and this has even been improved ten years later by a new data reduction almost single-handed by F.~Van Leeuwen  \cite{2007A&A...474..653V}. 

The publication gave the astronomical community a brand new astrometric catalogue of 120,000 stars, all accurate in position and parallaxes to about one thousand of a second of arc (two times better than the initial objective) and supplemented with photometric  \cite{1997A&A...323L..61V} and double star data  \cite{1997A&A...323L..53L}. On top of that came also a less accurate but much larger catalogue of 2.5 million stars called Tycho-2 \cite{2000A&A...355L..27H} resulting from measurements made with the Hipparcos star detector and the combination with the almost one-century-old photographic plates of the Carte du Ciel sky mapping.

  Concentrating on the main topic of this review, Hipparcos astrometry was a truly new start for parallax survey. The total number of trigonometric parallaxes rose at once to more than 100,000, with nearly 50,000 better than 20\% in fractional errors ($\sigma_\varpi/\varpi < 20\%$)  and 20,000 at the 10\% level. The reference frame was made inertial by linking the whole system to extragalactic sources, using radio stars common to Hipparcos and radio observations, or of observations of quasars relative to nearby Hipparcos stars \cite{1997A&A...323..620K}. This was an epoch-making advance in astrometry and in the measurement of stellar distances. Application to luminosity calibrations for a large variety of stellar types followed closely the publication and set the pace to improvements of the second rung of the distance ladder. More generally  the  Hipparcos data have influenced many areas of astronomy such as the the structure and evolution of stars and the kinematics of stars and stellar groups, the distance of the Hyades cluster, the galactic rotation from Cepheid variable stars,  albeit the limited sample size of sources and observed volume. The outstanding and in-depth review by Perryman  \cite{2012aaa..book.....P}  based on most of the papers published in 1997-2007 using the Hipparcos catalogue provided an amazing detailed survey of the application of Hipparcos to stellar and galactic physics. The stellar distances and accurate proper motions together with the high-precision multi-epoch photometry are the crucial data exploited in these papers.

 \subsection{The Gaia parallax survey}\label{sect:gaia}
    Hipparcos was a resounding and acclaimed international success allowing the Europeans to quickly submit several more ambitious proposals for space astrometry, at the same time as others were also submitted to NASA or to the Japanese space agency \cite{1996A&AS..116..579L}. Only one of these proposals survived the various examinations by selection committees and Gaia was eventually selected as a cornerstone mission in April 2000 for a launch around 2011. 
    
    The basic concept is directly drawn from Hipparcos, but with a much larger telescope (actually two telescopes), a mosaic of 106 CCD detectors replacing the outdated photoelectric detector of Hipparcos. Two other instruments were added to carry out spectrophotometry and spectroscopic measurements, the latter to measure the velocity along the line of sight. While Hipparcos catalogue was limited to 100,000 pre-defined stars brighter than 13.2 mag, Gaia was designed to realise a sensitivity-limited survey to 20 mag. Hipparcos could only take a star at a time while Gaia is able to record simultaneously several 10000s  images mapped on its focal plane. About one billion stars, amounting to $\approx 1$ percent of the
 Milky Way stellar content, are expected to be repeatedly observed during the nominal 5-year mission, with a final astrometric accuracy of $25$ {\muas} at $G = 15$ mag. (1 {\muas} = 0.001 mas = $10^{-6}$ arcsec). 
 
 Gaia's main scientific goal is to clarify the origin and history of our Galaxy, from a quantitative census of the stellar populations and extremely accurate astrometric measurements to derive proper motions and parallaxes. See \cite{2001A&A...369..339P} for the proposal and \cite{2016A&A...595A...1G} for a presentation of the actual mission, the spacecraft, the operations and the data acquisition strategy. The principle of the scanning satellite relies on a slowly spinning spacecraft to measure the crossing times of stellar images transiting on the focal plane. As for Hipparcos, there are two fields of view combined onto a single focal plane where astrometric measurements are done. The time relates the one-dimensional star position to the instrumental axes. The relation to the celestial frame is obtained with the satellite attitude, which is solved simultaneously with the star positions in a global solution as described technically by Lindegren et al. \cite{2012A&A...538A..78L}. 
 
 The Gaia satellite was launched on 19 December 2013 and the science data collection started after the in-flight qualification on 25 July 2014. A first batch of results was released on 15 September 2016 with only 14 months of data processed. This release comprised primarily a position catalogue (only two position parameters per source) for  1.14 billion stars, the largest ever such collection. A smaller catalogue combining Gaia and Hipparcos included parallaxes and proper motions for $\approx 2,000,000 $ stars with a sub-mas accuracy \cite{2016A&A...595A...2G}. The release contained also variable stars and a  set of 2200 quasars common to Gaia and the radio ICRF \cite{2016A&A...595A...5M}  used to align the Gaia and radio frames. Therefore the Gaia reference frame and ICRF are nominally identical.
   
\begin{figure}[h]
  \begin{center}
   \includegraphics[width=10cm]{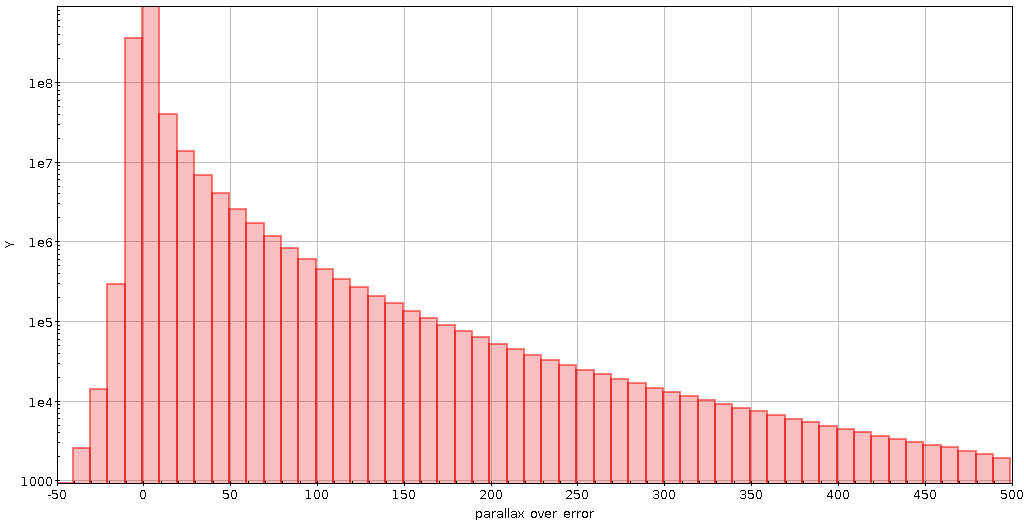}
  \end{center}
  \vspace{-5mm}
\caption{Distribution of the relative accuracy of the parallaxes in the Gaia survey, given as $\varpi/\sigma_\varpi$. There are about 50 millions stars with a parallax with a relative accuracy better than 10\%. Credit: ESA/Gaia/DPAC}
\label{Fig:sigma_varpi}
\end{figure}

  In April 2018 the 2nd release came out with parallaxes for nearly 1.4 billion stars \cite{2018A&A...616A...2L}, with a median uncertainty of 0.1 mas at G=17 and 0.7 mas at G = 20.  If we set the threshold for the usefulness of a distance to a relative precision better than $20\%$, at $G = 17$, Gaia DR2 reaches distances up to 2 kpc and  $0.3$ kpc at $G =20$. The distribution of the parallax fractional uncertainty is shown in Fig.~\ref{Fig:sigma_varpi} in the form of $\varpi/\sigma_\varpi$. One reads directly on the histogram that about 50 millions stars (out of 1.3 billion in the survey) have a relative precision in distance better than 10\%, to be compared to  $20,000$ with Hipparcos. This again will increase around to 100 million in the more advanced releases. 

  This makes up the state of the art today regarding our knowledge of the stellar distances within our Galaxy from the geometric measurement of their parallaxes. Gaia found that there are $620,000$ stars (more precisely unresolved stellar systems) within 100\,pc and nearly $5200$ at $d < 20$\,pc. This gives a typical density of $0.15$ star pc$^{-3}$ in the solar neighbourhood. For a uniform and random distribution with $\rho$ stars per cubic parsec, one computes easily the mean distance between two closest neighbouring stars as,
    \begin{equation}
      <d>  = \frac{4\pi \Gamma\left(\frac{4}{3}\right)}{3 \left(\frac{4\pi}{3}\right)^{4/3} \rho^{1/3}}\quad \simeq\quad \frac{0.554}{\rho^{1/3}}
    \end{equation}
 yielding with the Gaia census to $<d> \approx 1.04$ pc for the typical distance between a star and its nearest neighbour, not much different from the distance between the Sun and Proxima Centauri.

  Although this data release uses only 22 months of data, while the nominal mission will have 60 and a possible extension of almost five years is likely, this parallax survey is by far the most comprehensive ever done and has no match in terms of size and accuracy, with the exception, regarding the accuracy, of a handful of radio masers observed with the VLBI technique. For the Galactic stars this is the crowning of nearly two centuries of parallax quest starting with F.W. Bessel in 1838. Until many years in the future there will be not such undertaking to get trigonometric parallaxes directly from astrometric observations and the Gaia survey is now actively being (see Sect.~\ref{sect:cepheids}) used to reconstruct the whole distance scale beyond the Galaxy, based on secondary indicators. 
  
  I summarise in the next section the principles and ranges of applications of the numerous methods used today by astronomers to estimate distances from photometric indicators, but a major warning must be issued at this stage to any user deriving distances or other astrophysical parameters from the Gaia parallaxes.  Better to read first the paper by Luri et al. (\cite{2018A&A...616A...9L}) to guard against  the numerous pitfalls to get  a distance or a luminosity from the parallax. Only for the very accurate parallaxes ($\sigma_\varpi/\varpi < 0.1$) the straightforward transformation $ d = 1/\varpi$ can be used safely. Otherwise a better statistical inference must be used to keep the bias under control.
    
\section{Secondary indicators}\label{sect:secondary}
\subsection{Overall principles}
    I refer here to distance estimators which are not directly derived from the stellar parallaxes as described in the previous sections. Most of the methods are very simple in essence, since one tries to compare the apparent luminosity of a source, how much radiant power is received on the Earth, to the true luminosity of the same source. Most of the problems with these techniques come from  the selection of the standard candles,  their reliable identification, the calibration of the method, the systematic effect affecting them, the extinction of light during its journey to the Earth and the assumptions regarding the true luminosity of the sources. The determination of distances farther than the range accessible to trigonometric parallaxes follows more a less a single model with the following steps: 

\begin{itemize}
\item Identify a class of astronomical objects, bright enough to be seen at large distances
\item Prove that they have a well defined luminosity to qualify as standard candles
\item Measure the flux on the ground or from space around the Earth
\item Find their distances to calibrate their luminosity
\item Identify and select similar objects to find the distances of far-away galaxies
\item Calibrate a new rung of the ladder with these new distances
\end{itemize}

    Let $L$ be the absolute luminosity of an astronomical source, that is to say the total rate of luminous energy production, and $l$ the flux received on Earth per unit of surface. If we assume a propagation without loss of energy one has,
    \begin{equation}\label{eq:flux}
       l  = \frac{L}{4\pi d^2}
    \end{equation}
    where $d$ is the distance between the star and the Earth. If $l$ is measured and $L$ is known or estimated from the star physical properties, then one can estimate the distance. In practice luminosities are expressed in a magnitude scale, and the distances in pc are related to the difference between the apparent ($m$)  and absolute ($M$)  magnitude as,  
    \begin{align}\label{magnitudes}
      m  &= -2.5 \log l + C_1\\
      M  &= -2.5\log L + C_2
    \end{align}
    for the apparent and absolute magnitudes and with Eqs.~\ref{def_varpi} and \ref{eq:flux},
    \begin{equation}\label{m_minus_M}
      m - M = 5\log d -5= -5\log \varpi - 5   
    \end{equation}
 with the convention that the absolute magnitude is the apparent magnitude the star would have if placed at 10 pc. Here $\varpi$ is in arcsec, or equivalently the distance $ d = 1/\varpi$ is in pc. Classically the magnitude difference $ m - M$ is called the \textsl{distance modulus} of the source. While rarely used for stars in our Galaxy it makes sense for clusters of stars at several kpc or Mpc such as a globular cluster, a dwarf galaxy surrounding the  Milky way or a distant galaxy like Andromeda, as long as the source is resolved into stars. To illustrate this point the LMC (Large Magellanic Cloud), well visible in the southern sky, is located at about 50 kpc from the Milky Way. Therefore its distance modulus is $18.5$ mag and a star similar to the Sun ($M = 4.8$) would have an apparent magnitude of $23.3$, very faint for many of the telescopes in use today and not visible with Gaia. On the other hand, a classical Cepheid pulsating with a period of 4 days has $ M \approx -3$ and  would be seen as a star of $m = 15.5$ in the LMC, rather easy to detect with a medium size telescope and an accurate target for Gaia.
 
 The extinction along the path is probably the most serious issue near the galactic plane, which essentially amounts to saying that the  radiant flux decreases faster than  the  inverse square law. If one has an absorption coefficient of $\Gamma(l,b)$ in mag pc$^{-1}$  in the direction defined in galactic longitude $l$ and galactic latitude $b$, Eq.~\ref{m_minus_M} becomes for a source at distance $d$,

\begin{equation}\label{m_minus_M_gamma}
   m - M    = 5\log d -5 +\Gamma d
\end{equation}
 For stars the extinction comes with a reddening, since dust scatters more efficiently the shorter wavelengths and the spectrum appears redder than what is expected for a star with known spectral type and luminosity class. There is a rather well defined relationship between the reddening (called colour excess) allowing one to make the corrections from stars observed at the same place and in the same direction.  
 
 \begin{figure}[h]
  \begin{center}
   \includegraphics[width=10cm]{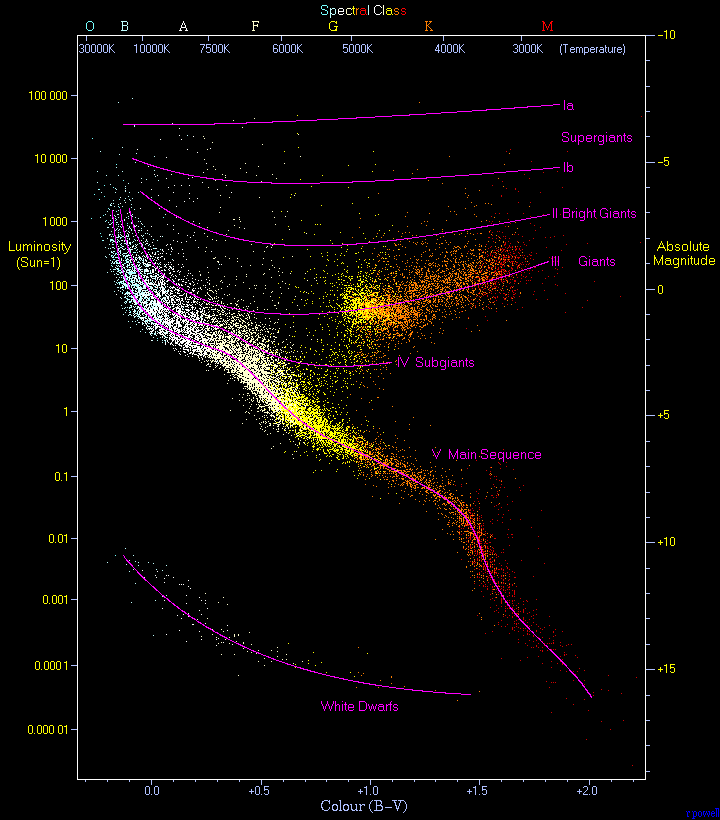}
  \end{center}
  \vspace{-5mm}
\caption{An observational Hertzsprung–Russell diagram with $22,000$ stars plotted from the Hipparcos Catalogue and $1,000$ stars from the Gliese Catalogue of nearby stars.} 
\label{Fig:HR_diagram}
\end{figure} 
 
\subsection{Distances of clusters}\label{sect:clusters}
This is the first level of photometric/spectroscopic-derived distance indicator relying on the absolute luminosity of stars and the technique providing the very first estimate of a distance for most of the stars in or out of the Galaxy. It is based on the location of stars on the Luminosity-colour diagram, usually referred to as the HR diagram named after E.~Hertzsprung and H.N.~Russel who discovered the feature independently around 1910. A diagram based on Hipparcos distances is shown in Fig.~\ref{Fig:HR_diagram}. Stars tend to fall only into certain regions of the diagram. The most prominent feature is the long concentration along a  diagonal crossing the diagram from the top left to the lower right. This is the location of the main sequence where most of the stars lie during their hydrogen burning phase. In the lower-left there is a narrow line with white dwarfs, and, above the main sequence,  several nearly horizontal lines with the subgiants, giants and supergiants, a state in the star life when core hydrogen is exhausted. The Sun is found on the main sequence  at $B - V = 0.66 $ and luminosity of $1 L_\odot$ by definition. The spectrum of a star and its location in the diagram are highly correlated, to the extent that a solar analogue can be recognised from the absorption lines of its spectra from the depth or absence of characteristic lines such as Hydrogen, Calcium, Oxygen etc. 

If a distant Sun is found from its spectral characteristics, one may say that its luminosity is similar to the Sun's and its absolute magnitude is close to $4.7$ is the V passband. Then confronted to its apparent magnitude a distance may be inferred with Eq.~\ref{m_minus_M}, if extinction can be neglected. Using the reddening, the extinction can be included with Eq.~\ref{m_minus_M_gamma} to get the distance as well. Due to intrinsic scatter between stars of similar properties, or because of different initial chemical composition, the presence of an unseen companion, this method is not very accurate when applied to individual stars, although it is useful to get a first estimate of the distance for remote stars. 

However the same principle becomes much more efficient when applied to a cluster of stars \cite{2009A&A...497..209V}, \cite{2017A&A...601A..19G}. Observing a cluster like the Hyades or the Pleiades one may identify stars that belong to the cluster from their distances, kinematical parameters (they should have the same space velocity), chemical composition and then exclude field stars not related to the cluster. These stars are thought to have been formed from the same initial cloud at the same epoch. They have the same age and share a similar content of heavy elements. Being similar and at the same distance, their distribution in the HR diagram displays narrow sequences, with little scatter, at least for all the single stars. The difference between the absolute HR diagram and that of a cluster using apparent magnitude is just a vertical translation equal to the cluster distance modulus since the $m-M$ is constant for all the members. By searching for the best fit between the main sequence of the cluster to a calibrated main sequence of the diagram one gets immediately the distance of the cluster. If extinction is significant, there is also a horizontal translation for the reddening. The calibration must be done beforehand  on the closest clusters, like the Hyades from the trigonometric parallaxes or the kinematics of the whole cluster combining the tangent and radial motions. There are many complications in the details (ages, metallicity, He abundance, that differ  from the reference cluster and displace the sequence), but the principles are as described and allow one to get distance estimates for most of the not too old open clusters found in the galactic plane where main sequence stars are visible. A solar-like star with $V \approx 4.7$ is brighter than $ m_v = 20$ up to $d =10$ kpc without extinction. So the method extends the distances achievable without trigonometric parallaxes to few kpc in the Milky-Way.

\subsection{Distances from the Cepheids}\label{sect:cepheids}
Using Cepheids as standard candle is the single most important distance indicator for extragalactic distances up to some 10s Mpc. Cepheids are supergiants stars of type F to K with regular brightness variation over periods ranging from 1 days to 50 days. The source of variability is well understood as an hydrodynamic instability causing the pulsation of the star which changes in size and surface temperature during the cycle. As supergiants Cepheids are intrinsically very bright and can be seen a very large distance, therefore detectable in external nearby galaxies. With a brightness of  $M \approx -5$,  a star seen with apparent magnitude $m=22$ is at a distance of (Eq.~\ref{m_minus_M}) 2.50 Mpc, three times as far as  the Andromeda galaxy. During the first decade of the 1900s H. Leavitt studied variable stars in the LMC and SMC (Large and Small Magellanic clouds, two companion galaxies at 50 kpc from the Milky Way), and found about 50 Cepheids in the LMC. She rightly noticed that the period of variability was all the more longer as the star was bright. Moreover she showed that the mathematical law relating the apparent magnitude and the logarithm of the period was linear \cite{1912HarCi.173....1L}. This was at once a major breakthrough in this field with far-reaching consequences for the understanding of the structure of the Universe. The early death of H. Leavitt deprived her of a likely Nobel Prize.

Given that  these stars were all at the same distance, one could infer that the same relation held for the absolute magnitude, and provided the link between the period and the luminosity (the Period-Luminosity or PL relation)  could be calibrated, one would know the distance of the host galaxy. Since then many calibrations have been published from census of galactic cepheids whose distances could be estimated by independent means.  They are relatively rare sources and their number  is  limited to few thousands, although many new have been discovered by Gaia. The population is rather uniform and the basic assumption is that Cepheids in external galaxies behave like those found in the Milky Way. The Gaia DR2 variability set comprises $9675$ stars classifieds as Cepheids, against only $599$, mainly in the region of the LMC, in the DR1 \cite{2017A&A...605A..79G}. This  represents  the first   full-sky census of Cepheids and provides a flavour of Gaia potential to  recover  most  of  the  Milky Way Cepheids \cite{2018arXiv180502079C}, not hidden by dust clouds. 
A typical P-L law has the form, with the period $P$ in days,
\begin{equation}\label{PL_01}
   M    = a  - b\log P
\end{equation}
or with a colour correction,
\begin{equation}\label{PL_02}
   M    = a  - b\log P + c(B-V)
\end{equation}
where the most important parameter is the zero point coefficient $a$. The coefficients $b$ and $c$ are independent of the distance and result from the analysis of the light curves. Other colour indices that $B-V$ are also  used. 

Until the advent of Gaia, the Cosmic distance scale rested primarily on the Cepheid calibration using the Hipparcos parallaxes with,
\begin{equation}\label{PL_TGAS}
M_V = -2.76\log P -1.45 
\end{equation}
with an estimated error in the range of $5-20$\%.  Mention also the HST derived calibration \cite{2007AJ....133.1810B}
\begin{equation}\label{PL_02}
 M_V    = - 3.34\log P + 2.45(V-I) -2.52
\end{equation}

\begin{figure}[h]
  \begin{center}
   \includegraphics[width=10cm]{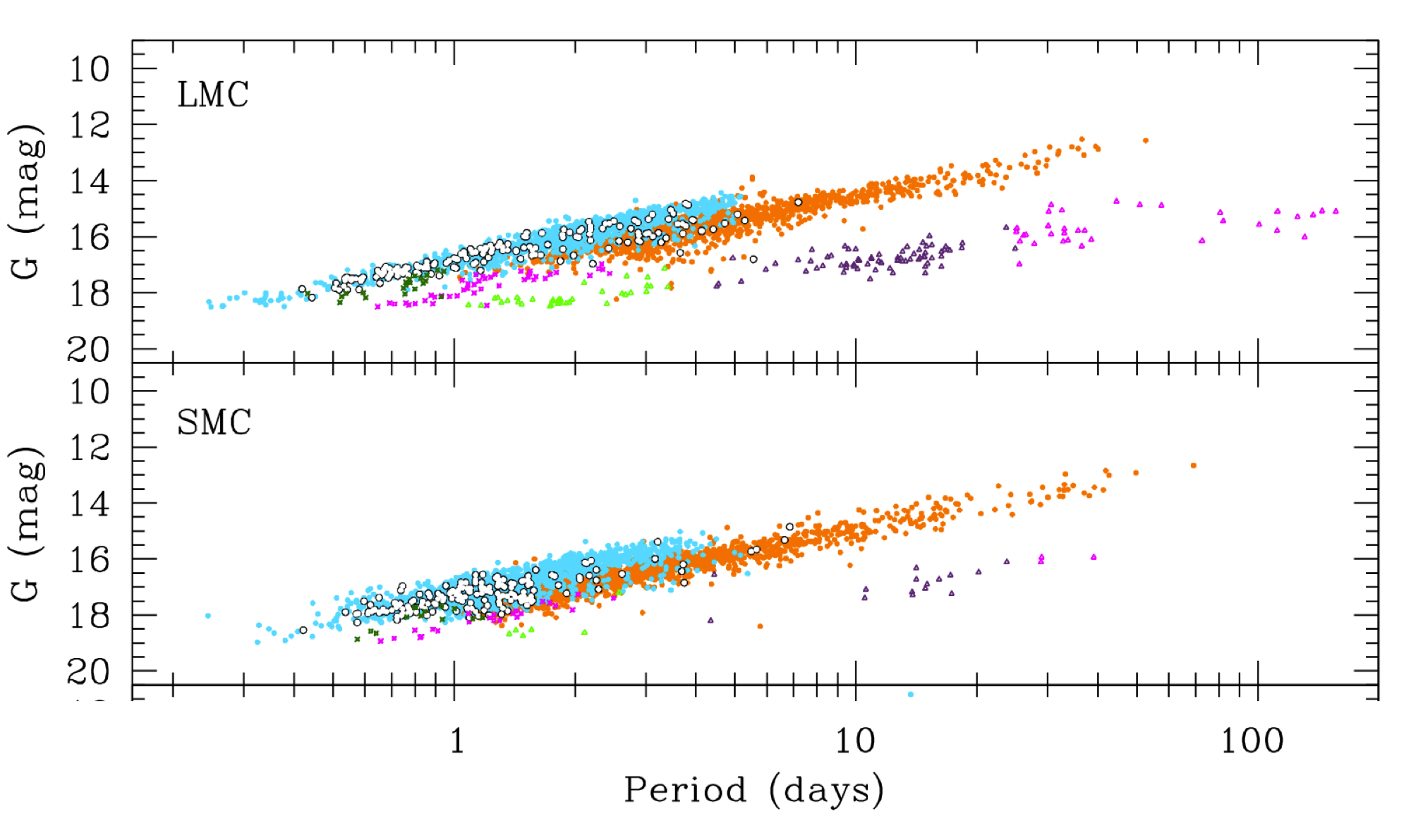}
  \end{center}
  \vspace{-5mm}
\caption{Period-Luminosity relation in apparent $G$ magnitude uncorrected from reddening in the LMC and SMC from the Gaia DR2 sample (adapted from \cite{2018arXiv180502079C}) }
\label{Fig:Cepheids_LMC_SMC}
\end{figure}

Using Gaia DR1 and distances from the TGAS solution (Gaia combined with Hipparcos and Tycho), Clementini and collaborators \cite{2017A&A...605A..79G} gave a new calibration for classical Cepheids in the $V$-band as,
\begin{equation}\label{PL_TGAS}
M_V = -2.678\log P - (1.54 \pm 0.10)
\end{equation}
See the paper for the details of the selection and the bias that may result. Taking the numbers as given one sees that a Cepheid with a period of 50 days, has a $M_V$ magnitude of $-6.1$ and will be visible with the HST and without extinction  at 2 Mpc. Similar calibrations have been also obtained for Population II Cepheids and the fainter, but much more frequent, RR Lyrae pulsating stars. The latter are much more common stars and are useful distance indicators for Globular clusters in the Milky Way halo up to 50 kpc.  Calibration of the Galactic Cepheids from the Gaia DR2 is not yet completed, but a partial result  for the Cepheids in the Magellanic clouds is shown in Fig.~\ref{Fig:Cepheids_LMC_SMC} from \cite{2018arXiv180502079C}.The colours code different types of Cepheids with slightly different Period-Luminosity relations. The plots are impressive by themselves for the number of sources, the resolution between the different populations and the small scatter around a visual linear fit. As the LMC/SMC distance modulus can be obtained by independent techniques, the curve can be transformed into absolute calibrations (with correction for the reddening) and compared to the Galactic Cepheids.

\subsection{Towards cosmological distances}
As detected so far, the spiral galaxy NGC 3370  contains the farthest Cepheids yet found at a distance of 29 Mpc. To reach distances where the Hubble flow becomes predominant other rungs are required for galaxies beyond $500$ Mpc. So far the SNe~Ia are the most relevant sources to be used as standard candles for the very large distances. SNe Ia result from the catastrophic instability of a binary white dwarf accreting material from its companion star and exploding when it reaches the Chandrasekahr limit. This well defined particular condition accounts for the relative uniformity of the observable properties, such as the light curve of the SN Ia and their maximum brightness. They are recognised from other SNs by the shape of their light curve after the maximum, their spectra and they constitute good standard candles with the peak luminosity $M_V \approx -19.5$, corresponding in energy output to about $10^9 L_\odot$ \cite{1999ASPC..167..217M}. They are seen in all types of galaxies with typically one event per galaxy every five centuries. Using Eq.~\ref{m_minus_M}, one sees that with the HST they can be seen at few Gpc distances, that is to say at the start of the cosmological distances. But this peak, standardised for different light curves, needs to be calibrated and again Cepheids are used for this purpose within galaxies at rather small distances of few Mpc  as explained by Sandage and Tammann \cite{2006ApJ...653..843S} in  a classical paper. 

A very important application of the Gaia DR2 Cepheids dealing with this topic has been reported in \cite{2018ApJ...861..126R} with the combination of the HST photometry of 50 Cepheids located in galaxies at $d < 50$ Mpc where Supernovae Ia have been found and used to extend the distance scale to Gpc and constrain the Hubble constant. Basically this fills the necessary step to assess the absolute luminosity of SNe~Ia  within relatively nearby galaxies from a distance estimate of these galaxies based on another standard candle. Gaia Cepheids in the MW are the most coveted source to achieve this goal given their brightness ($G < 12$) and then their expected high  parallax accuracy, about five times better than the HST astrometry. From the HST data and a previous Cepheid calibration using $H_0 = 73.24$ km\,s$^{-1}$\,Mpc$^{-1}$ the authors  have calculated the absolute magnitude of the Cepheids in the HST photometric system with the $P-L$ relation,
\begin{equation}\label{PL_TGAS}
M_H = -5.93 - 3.26(\log P-1)  
\end{equation}
The analysis done in \cite{2018ApJ...861..126R} confirms the existence of a bias in Gaia DR2 parallaxes, but larger than the Gaia quoted value of $-29$ {\muas} based on fainter quasars \cite{2018A&A...616A...2L}. In the magnitude range of bright Cepheids they found $ -46 \pm 6$ {\muas} instead. This has an implication for the Hubble constant, since the HST value is not in agreement with that determined from Planck cosmic microwave background (CMB) data which yields $H_0 = 66.93 \pm 0.62 $ km\,s$^{-1}$\,Mpc$^{-1}$. However no such tension appears in \cite{2018arXiv181002595S} if we extend the DR2 bias found for the quasars to the bright Cepheids and the Planck value of $H_0$. The issue is not solved yet but is just mentioned to show that even with the best tools in the hands of astronomer, as Gaia and HST are, the metrology is never simple and extreme care must be exercised everywhere. With Gaia parallaxes and their sheer number, a new page just opens up and new papers are expected in the coming years discussing and questioning the cosmic distance scale established with different techniques.

\section{Conclusion}
Large distances are the realm of astronomers with the characteristic that one cannot experiment but only deal with the information we can collect from the Heavens, primarily of electromagnetic nature, even though new promising messengers such as the neutrino astronomy and the emerging one with gravitational waves,  are just ahead of us with new challenges. In this brief overview, given the broadness of the field, I have attempted to show that an astronomer dealing with precision measurements must display the same rigour as a metrologist in his/her laboratory, by ceaselessly controlling his apparatus and above all calibrate and calibrate again.  The metrological spirit pervades every area of experimental and observational science, whatever the scale of space or time under consideration.

 
\section*{References}

\bibliography{FM_DS_bibfile,BiblioICRF}

\end{document}